\documentclass[12pt,preprint]{aastex}

\shorttitle{supernova n120}

\begin{document} \par

\title{The XMM-Newton X-ray emission of the SNR N120 in the LMC }

\author{Jorge Reyes-Iturbide\altaffilmark{1}}
\affil{Instituto de Astronom\'\i a,
Universidad Nacional Aut\'onoma de M\'exico, Apdo. Postal 70-264, C.P. 04510, M\'exico, D. F., M\'exico}
\email{jreyes@astroscu.unam.mx}
\author{Margarita Rosado\altaffilmark{1}}
\affil{Instituto de Astronom\'\i a,
Universidad Nacional Aut\'onoma de M\'exico, Apdo. Postal 70-264, C. P. 04510, M\'exico, D. F., M\'exico}
\email{margarit@astroscu.unam.mx}
\author{Pablo F. Vel\'azquez\altaffilmark{2}}
\affil{Instituto de Ciencias Nucleares, Universidad Nacional Aut\'onoma de
  M\'exico, Ap. Postal 70-543, C. P. 04510, M\'exico D.F., M\'exico}
\email{pablo@nucleares.unam.mx}

\begin{abstract}
We present new XMM-Newton observations of the supernova remnant N~120
in the LMC, and numerical simulations on the evolution of this
supernova remnant which we compare with the X-ray
observations. The supernova remnant N~120, together with several
HII regions,  forms a large nebular complex (also called N~120) whose shape
resembles a semicircular ring. From the XMM-Newton data we generate
images and spectra of this remnant in the energy band between 0.2 to
2.0~keV. The images show that the X-ray emission is brighter towards
the east (i.e., towards the rim of the large nebular complex). The
EPIC/MOS1 and MOS2 data reveal a thermal spectrum in soft X-rays. 2D
axisymmetric numerical simulations with the Yguaz\'u-a code were
carried out assuming that the remnant is expanding into an
inhomogeneous ISM with an exponential density gradient and showing
that thermal conduction effects are negligible.  Simulated X-ray
emission maps were obtained from the numerical simulations in order to
compare them with the observations. We find  good
agreement between the XMM-Newton data, previous optical kinematic
data, and the numerical simulations; the simulations reproduce the
observed X-ray luminosity and surface brightness distribution. We have also
detected more extended diffuse X-ray emission probably due to the N~120 large
HII complex or superbubble.

\end{abstract}


\section{Introduction}

X-ray emission from supernova remnants (SNRs), bubbles, and superbubbles 
provides an important means for the study of how the stars interact with the 
interstellar medium (ISM) and deliver part of their energy and matter to 
the galaxy that hosts them.  It also gives important clues on the particular 
mechanisms involved such as supernova explosions and stellar winds.  In the 
case of SNRs, the X-rays emitted allow us to diagnose their nature, to 
determine the energy released by the supernova explosion,  to gain insight 
into the density distribution prior to the explosion, and to estimate the shock 
velocity and dynamical age, among other issues. It is also important to confront X-ray, optical, and radio
observations  with numerical simulations of SNR and bubble evolution,  to discriminate the  main characteristics of SNR 
evolution such as density gradients, thermal conductivity and magnetic fields, 
among others. This idea  could be applied to more complex situations such as 
multiple SNR evolution, supernova explosions within wind-driven bubbles, or 
interactions between bubbles and SNRs, which we intend to explore in the future. 

Accounting for all these motivations, we used XMM-Newton data of 
SNR N~120 in the Large Magellanic Cloud (LMC) to compare the X-ray 
luminosity and surface brightness distribution with results from axisymmetric numerical simulations obtained with the 
adaptive grid Yguaz\'u-a code (see, for example, Vel\'azquez et al. 2001). We 
selected this SNR because its kinematics (at optical wavelengths) has been 
already studied, having shown that the SNR is well differentiated from other 
neighboring nebulosities. An advantage of studying the X-rays from an object 
in the LMC is that the interstellar reddening towards this galaxy is small, 
thus making the LMC objects ideal to be studied at X-ray wavelengths. Furthermore, 
the distances to Galactic SNRs are not easy to determine while in the LMC all the objects are at the same (known) distance. 
Therefore, by selecting an SNR in the LMC (with a well known distance derived from 
several independent ways) we can be quite sure of its estimated linear 
diameter. 

The supernova remnant N~120 (0519-697) was first identified as a SNR by 
Mathewson \& Clarke (1973) through its non-thermal radio spectrum and its
 high [SII]/H$\alpha$ line-ratio. Its linear diameter is 30 pc for an LMC 
distance of 50 kpc (Feast 1999). Dickel \& Milne (1998) obtained high spatial resolution radio-continuum maps at 1380 and 2378 MHz, showing a  shell with a brightness enhancement towards the east, where the shell is well defined. They also confirm the non-thermal nature of the radio emission and obtain a spectral index $\alpha$ = --0.47. 
The X-ray emission from SNR~N120 has been 
studied with Einstein data by Mathewson et al. (1983), who show that this
SNR is one of the faintest X-ray sources among the known SNRs in the LMC
(its derived X-ray luminosity is about 10$^{35}$ ergs$^{-1}$). SNR N~120 is located within the 9$\arcmin$ $\times$ 7$\arcmin$ 
(or 135 $\times$ 105 pc) arc-shaped nebular complex N~120 
(Henize 1956) or DEM 134 (Davies et al. 1976), which consists of
small wind-blown bubbles, HII regions, and the SNR N~120 
along the rim at a common systemic velocity (Laval et al. 1992). Within this superbubble there has recently been detected OVI ($\lambda$1032 \AA~) emission  by FUSE (Sankrit \& Dixon 2007) suggesting that shocked hot gas is present.

At optical wavelengths (H$\alpha$ and [SII]), SNR N~120 is seen as 
an ellipsoidal shell. The  [SII] emission, as well as the highly receding H$\alpha$ emission 
show a shell whose eastern side is brighter than its western 
side (Georgelin et al. 1983, Rosado et al. 1993). The kinematics of SNR N~120 have been 
studied in detail by Rosado et al. (1993) by means of photometrically 
calibrated H$\alpha$ scanning Fabry-Perot observations. Rosado et al. (1993) also derive the following 
parameters for the SNR assuming that the optical emission is due to shocks 
induced by the primary blast wave in dense clumps: clump density, 
n$_c$ = 8 cm$^{-3}$, intercloud density, n$_b$ = 0.1 cm$^{-3}$, velocity 
of the shock induced in the clumps = 100 km s$^{-1}$, and shock velocity
in the intercloud medium = 800 km s$^{-1}$. With these values, they 
computed a postshock temperature of 8.7 $\times$ 10$^6$ K, an age of 
7300 yr, and a released energy of 5 $\times$10$^{50}$ ergs.

In Section 2 we present the data and data reduction. 
Section 3 is dedicated to the presentation of the observational results 
and a discussion of their implications. In Section 4 we present the 
numerical code, the model and the results of our numerical simulations 
on the particular conditions found for SNR N~120, and in Section 5 
we give the conclusions.

%% In a manner similar to \objectname authors can provide links to dataset
%% hosted at participating data centers via the \dataset{} command.  The
%% second curly bracket argument is printed in the text while the first
%% parentheses argument serves as the valid data set identifier.  Large
%% lists of data set are best provided in a table (see Table 3 for an example).
%% Valid data set identifiers should be obtained from the data center that
%% is currently hosting the data.
%%
%% Note that AASTeX interprets everything between the curly braces in the 
%% macro as regular text, so any special characters, e.g. "#" or "_," must be 
%% preceded by a backslash. Otherwise, you will get a LaTeX error when you 
%% compile your manuscript.  Special characters do not 
%% need to be escaped in the optional, square-bracket argument.

\section{XMM-Newton  Data and Data Reduction}

SNR N~120 was observed with the XMM-Newton satellite (Jansen et al.\ 2001). The data were obtained during Revolution 0337 in October 
2001 using the EPIC (European Photon Imaging Camera) which consists of 
two MOS  (Metal Oxide Semi-conductor) CCD arrays (MOS1 \& MOS2) (Turner et al. 2001), thus EPIC/MOS1 and EPIC/MOS2 
cameras are used in this work (Obs ID 0089210701). EPIC cameras are CCD 
detectors with spectral resolution E/$\Delta$E $\sim$ 20-50 in the 0.1-10 keV 
band. The pointing coordinates were $\alpha$ (2000) = 05h18m42.0s, 
$\delta$ (2000) = --69$\arcdeg$39$\arcmin$30$\arcsec$. The two EPIC/MOS 
cameras were operated in the Full-Frame Mode, using the full 
30$\arcmin$ field of view of XMM-Newton, covering the entire dimensions
of SNR N~120 and including the nebular complex N~120, for a total exposure time of 62 ks.

For all observations, the Medium Filter was used. The Medium Filter is an aluminized optical blocking filter aimed at reducing the IR, visible and UV photons to which the EPIC/MOS cameras are also sensitive and whose detection would preclude the X-ray analysis. It is made of 1600 \AA ~of poly-imide film with 800 \AA ~of aluminium evaporated onto one side. This filter was used to prevent ``optical loading'' of the CCD, which can distort the X-ray spectra.
The XMM-Newton pipeline products were processed using the  XMM-Newton Science Analysis Software (SAS version 6.1.0).  The data were taken in two intervals, hereafter referred to as ``scheduled'' and ``unscheduled''. The difference between these two types of files arose because in the course of exposing the scheduled observations, there were interruptions due to high radiation levels;
the exposures taken after the interruptions we call  unscheduled.

The EPIC/MOS event files were screened to eliminate events due to charged
particles or those associated with periods of high background; only events
with CCD patterns 0-12 were selected. We discarded data with a high background
(i.e., count rates $\ge$ 1.0 and 1.0 counts s$^{-1}$ for the EPIC/MOS1 and
EPIC/MOS2 scheduled data, respectively, and 1.3 and 1.2 counts s$^{-1}$ for
the EPIC/MOS1 and EPIC/MOS2 unscheduled data, in the background-dominated
10-12 keV energy range). The resulting exposure times are 13.5 and 13.6 ks for
MOS1 and MOS2 scheduled, and 35.0 and 34.3 ks for MOS1 and MOS2 unscheduled
data.

To obtain an image of the X-ray emission of SNR N~120,  we merged  the screened event files of all the  EPIC observations.  We extracted an EPIC image in the 0.2 - 2.0  keV band with a pixel size of 4$\arcsec$.35 and a PSF of 9$\arcsec$ estimated from point sources in the same field.  

However, for the fitting of the X-ray spectrum of SNR N~120, we preferred not
to merge the different data. This is because the exposure times were different
and also because the task MERGE is not recommended for quantitative analysis.
For the analysis of the X-ray spectrum  the four individual data sets
were simultaneously fitting using XSPEC. In this way, we extracted spectra from each of the
four EPIC/MOS1 and MOS2 event files, scheduled and unscheduled, using a
circular aperture of 80$\arcsec$ radius, large enough to include all the X-ray
emission from the SNR.  We determined the background level outside the source
by extracting the spectrum in another circular aperture of similar radius to
that of the source, well outside the SNR emission and without encompassing
other X-ray sources.  The background spectra were then scaled and subtracted
from the source spectra.  The four resulting spectra were analyzed jointly
using the XSPEC spectral fitting package.  We also extracted a spectrum from
the merged event file, and analyzed this ``merged'' spectrum similarly, as a
matter of comparison.

%% In this section, we use  the \subsection command to set off
%% a subsection.  \footnote is used to insert a footnote to the text.


%% This section also includes several of the displayed math environments
%% mentioned in the Author Guide.

\section{Results: X-ray Brightness Distribution, Luminosity and Spectrum of the SNR N~120}

\subsection{ X-ray brightness distribution} 
%\label{bozomath}

Figure 1 shows the merged  EPIC image in the 0.2--2.0 keV band (field-of-view =495$\arcsec$ $\times$354$\arcsec$), in false color, showing the X-ray emission of the SNR N~120, with contours of the same emission superposed.  An image in the 2.0--5.0
keV band that we have produced (not shown) reveals no appreciable X-ray 
emission, and is very noisy. 

As seen in  Figure 1, the X-ray emission from the SNR (centered at $\alpha$ (2000) = 05h18m42.0s, $\delta$ (2000) = --69$\arcdeg$39$\arcmin$30$\arcsec$) shows an elliptical patch whose major
axis is approximately oriented along the E-W direction, with a maximum elongation of 140$\arcsec$.
 The eastern side of this emission is much brighter than the western side, particularly at the eastern and southeastern boundaries. Figure 2 shows the three-color optical image (from the Magellanic Clouds Emission-Line Survey, MCELS) of the SNR~120 in the H$\alpha$ (red), [SII] (green), and [OIII] (blue) with some of the isocontours of X-ray emission depicted in Fig.1 overlayed for comparison. Inspecting this figure, we see that at optical wavelengths SNR N~120 appears as an elliptical shell of 108$\arcsec$ $\times$ 82$\arcsec$ (27 pc $\times$ 20 pc), 
whose major axis runs along the NE-SW direction. The bright eastern and southeastern X-ray rims correlate 
quite well with the limb-brightened optical [SII] emission at those locations (as well as with H$\alpha$ emission at high receding velocities; see Rosado et al.'s 1993 Figure 2). However, while the optical emission also appears  ellipsoidal, its  major axis is almost perpendicular to the major axis of the X-ray emission.
 The dimensions of the X-ray source 
(taken from the 3rd contour corresponding to a flux 3$\sigma$ above the background level)
are 141$\arcsec$ $\times$ 127$\arcsec$ (equivalent to 34 pc and 31 pc 
at the LMC distance).

A close inspection of Figure 1 suggested that the bright X-ray emission of SNR N~120 is embedded into faint, diffuse emission, well outside the boundaries of SNR~120. To gain insight regarding  this diffuse and faint X-ray emission,  we  increased the signal to noise ratio by spatially smoothing  the EPIC image shown in Figure 1 with a Gaussian function of 2.5 pixels  width giving a scale of 11$\arcsec$ instead of the original scale of 4$\arcsec$.35 per pixel. Since the faint emission is diffuse, such a loss in angular resolution is not so important. The smoothed image is shown in Figure~3, also in false color and with contours of the emission superposed.  
From Figure 3 we see that, indeed, some diffuse X-ray emission is detected at faint levels (1st contour corresponding to a flux of 4 $\sigma$ above the background level) while SNR N~120 is the brightest source in the field.

Figure 4 shows the surface brightness contours from Figure 3 (specially the 1st contour mapping the diffuse X-rays) 
superposed onto a larger field of view three-color image (from the MCELS) of the N~120 HII complex containing SNR N~120 in its NW boundary, in H$\alpha$ (red), 
[SII] (green), and [OIII] (blue). Note that the optical 
counterpart of SNR N~120 has the highest [SII]/H$\alpha$ ratio of all the 
nebulosities of the N~120 HII complex, as revealed from the colors in the MCELS 
image, and that the eastern rim of the shell has a higher  [SII]/H$\alpha$ ratio than the western rim.

The faint, diffuse X-ray emission seems to correlate
with some of the nebulae
forming the N~120 complex (also called N~120 superbubble), all of them with thermal radio continuum emission
(see Laval et al. 1992 for further details on their exact location and Dickel \& Milne 1998 for further details about their radio spectral index).  These nebulae are N~120B,
N~120C1 and N~120C3 (to the East), part of N~120D (to the SW), part of the NW
arc  (N~140), and also covers part of the interior of the N~120 HII complex or superbubble. As mentioned in the Introduction, hot gas emitting in the UV has been detected from the N~120 superbubble (Sankrit \& Dixon 2007) and, consequently, the diffuse, low intensity X-ray emission could come from the N~120 superbubble.
The low intensity of the diffuse emission precludes obtaining insight into its
origin from the available data. The diffuse X-ray emission could arise from hot
gas escaping from the SNR N~120 or it could be associated with the N~120
superbubble, among other possibilities.
However, the tenuous X-ray emission follows so well the overall morphology
of the N~120 HII complex that we suggest that this faint emission could be associated with the N~120 superbubble and, therefore, not related to the SNR emission.

Figure 5  shows the contours of  X-ray 
emission displayed in Figure 1 (i.e., at the original resolution 4$\arcsec$.35 per pixel) superposed onto the  contours of the radio-continuum emission studied by Dickel \& Milne (1998). The spatial resolution of the X-ray and radio-continuum contours is similar. 
Both emissions are seen elongated in the same direction and have, approximately the same extension. The brightness enhancement towards the eastern boundary of the SNR seen in X-rays coincides with a similar enhancement in the radio-continuum emission. 
 
A previous X-ray image of SNR N~120 using ROSAT data was shown by Williams et al. (1999).
Our image (Figures 1, 2 and 3) obtained with XMM-Newton data shows similar features: a bright ellipsoidal patch embedded in fainter emission.  In both images, the brightest X-ray emission is located  superposed onto the eastern rim of the optical emission. However the direct comparison with multiwavelength data (optical and radio) allowed us to have a more realistic interpretation of the X-ray data.  

To gain further insight into the soft X-ray emission of SNR N~120 we extracted
a brightness distribution profile along the maximum elongation. Figure 6
(left) shows the mean aperture location (the width of the aperture is
$19\arcsec $ (5 pc) and the length is 174$\arcsec$ (42 pc)) superposed onto the X-ray image and (right) the
brightness distribution of the X-ray emission along this aperture. Several
positions are indicated to facilitate the identification of features. From the
brightness distribution profile, one can see that the maximum emission comes
from the eastern region which coincides in location with the bright [SII] rim
of the optical SNR. An irregular decrease in brightness follows towards the
western boundary of the ellipsoidal patch. The irregularities in brightness
found in the interior may be related to blobs of X-ray emission, which are
less bright than the emission at the boundaries of the SNR.

\subsection{X-ray spectrum and luminosity} 
%\label{bozomath}

For spectral analysis, we used XSPEC, version 11.3.1 (Arnaud 1996). The
spectra extracted from the EPIC/MOS1 and EPIC/MOS2 event files (as discussed
in Section 2) are given in Figure 7. This figure shows that the spectrum is
quite noisy for X-rays harder than about 2.0 keV. Therefore, we proceeded to
model only the softer part of the spectrum having significant signal (up to
2.0 keV). In order to get the physical conditions of the X-ray emitting gas,
we simultaneously modeled the MOS1 and MOS2 spectra by fitting different
spectral models such as: 1) absorbed MEKAL optically-thin plasma emission
models (Kaastra \& Mewe 1993; Liedhal et al.\ 1995; Morrison \& McCammon
1983), 2) absorbed PSHOCK models (Borkowski et al. 2001), and 3) absorbed
non-equilibrium ionization, NEI, models (Borkowski et al. 2001).  We froze the
chemical abundances to the average value of the interstellar medium in the LMC
(i.e., 0.3 times the solar abundances; Russel \& Dopita 1992; Hughes et al.
1998), and we used reasonable values for the absorption column density, N$_H$,
which should be in agreement with the measures of column densities in the LMC
direction (Dickey \& Lockman 1990), specifically, N$_H$ $\simeq$ 1$\times$
10$^{20}$cm$^{-2}$. In doing so, all considered models reproduce the observed
spectra with reasonable $\chi^2$ statistics. Table 1 lists the parameters of
the different fits to the X-ray spectra. It is interesting to note that PSHOCK
and NEI models predict higher plasma temperatures.  In view of the degeneracy
of X-ray emission models with respect to the X-ray spectra, we selected the
simplest emission model as representative of the X-ray spectrum of SNR N~120,
that is, the one with the emission of a thermal optically-thin plasma in
ionization equilibrium. It is possible that the plasma is not in ionization
equilibrium, but the data we have do not allow us to discriminate this issue.
In Figure 7 we reproduce (right panel) the plot of N$_H$ versus kT for the
confidence levels of the $\chi^2$ fit. Also displayed in Figure 7 is the
best-fit model over-plotted on the EPIC spectra.  In particular, this fit has
an absorption column density of N$_H$ = 1.21$\pm$0.66$\times$
10$^{20}$cm$^{-2}$, in agreement with the measures of column densities in the
LMC direction (Dickey \& Lockman 1990), a plasma temperature of T = 6.0
$\times$ 10$^6$ K (or kT = 0.31$\pm$0.06 keV ), and typical LMC abundances
(see Table 2). The absorption-corrected X-ray luminosity in the 0.2--2.0 keV
band is 1.2$\times$10$^{35}$ ergs $s^{-1}$ and flux 4.17$\times$ 10$^{-13}$
ergs cm$^{2}$ s$^{-1}$, confirming previous claims that SNR N 120 is one
  of the faintest SNRs detected in the LMC.

\section{Numerical Simulations}

\subsection{The model}

At radio frequencies, the SNR N~120 looks like an elongated shell with maximum elongation along the east-west direction. An increase of the radio continuum emission is observed to the east. This strong radio emission is correlated with strong [SII] emission (see Figures \ref{fig2} and \ref{fig5}). 
As mentioned in Section 3, the X-ray brightness distribution also  increases  towards the east.

The brightness asymmetry  could be produced 
by the collision of the SNR with high-density regions in the SNR environment, 
or by a density gradient of the ISM where the remnant is expanding.  At high density sites, the emission in both radio and X-rays is enhanced. 
Around SNR N~120, there are several HII regions (N~120A, N~120B, N~120C 
and N~120D; see Laval et al. 1992) which, together with the SNR, comprise the N~120
superbubble. For these reasons, we have modeled the 
ISM around this remnant as a medium with  increasing density to the east.  Hnatyk \& Petruk (1999) and Vel{\'a}zquez et al. (2004)
studied the evolution of SNRs in this kind of ISM and found that radio continuum and X-ray morphologies, such as those observed in SNR N~120, can be reproduced by choosing an exponential density profile for the surrounding environment. In their study, Hnatyk \& Petruk (1999) considered the importance of including projection effects, because the remnant could give the appearance of shell-like morphology  at radio frequencies, while the morphology is a filled center in X-rays, i.e., the remnant belongs to the mixed-morphology or thermal-composite group, which could result from  projection effects.

In the last 10 years, several theoretical works (Schneiter, de la
Fuente \& Vel{\'a}zquez 2006, Tilley \& Balsara 2006, Vel{\'a}zquez et
al. 2004, Cox et al. 1999, Shelton et al. 1999) analyzed the influence
of thermal conduction on the morphology of SNRs, focusing on the case
of the mixed-morphology SNR group. These works consider SNRs as
expanding into dense environments (with numerical densities of the
order of 10 cm$^{-3}$), producing a fast evolution of the
remnant. The effects of thermal conduction on the SNR morphology in
both radio and X-ray emission become important for more evolved SNRs,
i.e., the remnants that have entered into the third phase of their
evolution, the radiative phase. In this case, the remnant has a filled
center in the X-ray band, while it is a shell-type in radio
emission. Rosado et al. (1993) determine that SNR N~120 is in the
second phase of evolution, also called the Sedov phase, and calculate
an age for this remnant of 7300 yr. For these reasons, it seems that
the effects of thermal conduction can be neglected for the SNR N~120
case. Nevertheless, we have also explored the effects of thermal
conduction, as discussed below.

Based on these hypothesis, we have carried out numerical simulations modeling an SNR that evolves in a non-uniform ISM, with a density gradient in the form of an exponential law (this implies that there is a preferential direction).
Because of this main characteristic of the ISM, we employ the 2D
axisymmetric version of the Yguaz\'u-a code (Raga et al. 2000 and also Raga et al.
2002), which has a binary adaptive grid. The symmetry axis is in the density gradient direction.
The gas dynamic equations are integrated with a second-order accurate 
implementation of the flux vector splitting method of Van de Leer (1982) 
together with rate equations for the following atomic/ionic species: 
HI, HII, OI, OII, OIII, OIV, CII, CIII, CIV, HeI, HeII, HeIII, SII, SIII,
NI, NII, and NIII, which determine the radiative cooling function.

\subsection{Initial conditions}

We have made three  numerical simulations, employing a four-level binary 
adaptive grid in a 75$\times$37.5~pc (axial$\times$ radial) computational 
domain. The maximum resolution was  2.2$\times 10^{17}$~cm.

In the simulations, the SNR  explodes into an ISM with an exponential density profile given by
$n=n_{0}e^{-x/H}$, where $H$ is a characteristic length-scale, and n$_{0}$ is the density of the ISM at the position where the SN explosion occurs.

The SNR was located on the symmetry axis at the position 
$x_{SNR} = 7.5\times 10^{19}$~cm or 25 pc. The SN explosion is simulated by 
imparting a sudden release of energy, $E_0$, to the medium, independent of the 
explosion  mechanism and evolution of the pre-SN stage of the progenitor star. 
The released energy is put  into an expanding 5 $M_{\odot}$ sphere of constant 
density and 1 pc  radius which starts its interaction with the surrounding 
medium, simulating the SN explosion. Although this way of including the 
explosion does not give a good description of the early SNR evolution, for a SNR 
in the adiabatic phase or older (where $M_{ejected}<< M_{swept-up}$), the 
simulated explosion approaches in its effect an actual SN explosion. Incidentally, 
this approach has some basis if one considers that empirical mass-loss rates of solar 
metalicity stars with $ M \geq 35 M_{\odot} $ are thought to have pre-SN stages as 
hydrogen-free stars of about 5$M_{\odot}$ ( Woosley et al. 2002).

Based on Rosado et 
al. (1993) we chose an initial SN energy explosion, E$_0$, of $5\times 10^{50}$~erg, 
and an average ISM density (at the SNR position) of 0.1 cm$^{-3}$.
The scale length, H, was set as 10 pc, 5 pc, and 3.67 pc for models 1, 2, and 3, respectively.

\subsection{Simulated X-ray emission}

Synthetic X-ray emission maps were generated from numerical simulations, in
order to do a direct comparison with XMM-Newton observations. In all these
simulations Ionization Equilibrium was assumed due to the following reasons.
The observed spectra were fit employing different models, which consider
Ionization Equilibrium (IEQ) and Non-equilibrium Ionization (NEI). We use the
models MEKAL (IEQ) or pshock (NEI) and NEI (NEI). Simulated X-ray emission
that considers NEI models, is based on 1D calculations (by analytical laws or
1D hydrodynamical simulations). These models fit well the observed spectra
(see table 1) but give a partial hydrodynamical description of the SNR
evolution. Since SNR N~120 exhibits a pronounced asymmetry, it implies that 1D
models do not give a complete hydrodynamical description of this remnant. In
view that our main goal is to reproduce the observed surface brightness
distribution, and given that both IEQ and NEI models fit equally well the
observed spectrum, for generating the simulated X-ray emission we considered
IEQ models because they are simpler to implement and combine with our 2D
numerical results. The main difference between NEI and IEQ models is that at
low energies ($<2$ keV), IEQ models predict fluxes lower than the obtained by
NEI (Hughes \& Helfand 1985).

Thus, we used the CHIANTI database (Dere et al. 1997, Landi et al. 2006) for
calculating the X-ray emission coefficient $j_{\nu}(n,T)$ in the energy range
0.15-2.0 keV. We consider an average abundance for the Large Magellanic Cloud
of 0.3$Z_{\odot}$ and a column density $N_{HI}=1.2\times 10^{20}\ 
\rm{cm}^{-2}$, in agreement with the fit of the observed X-ray spectrum (see
Section 3.2). We also consider the IEQ as given by the Mazzota et al. (1998)
model. The $j_{\nu}(n,T)$ coefficients are calculated in the limit of low
density resulting $j_{\nu}(n,T)\propto n^2$.

\subsection{ Theoretical results}

In all models, the numerical simulations were carried out up to an integration
time of 8000 yr (about the estimated age of SNR N~120 given by Rosado et al. 1993).
At this time, the vertical diameter is of the order of 30 pc, in agreement
with our X-ray observations. Also, we obtained from our numerical results a
shock wave velocity of the order of $\mathrm 800\ km\ s^{-1}$, which
coincides with the value given by the kinematical study of Rosado et al. (1993).

Figure 8 compares the X-ray emission (in gray levels) obtained from models 1,
2 and 3 (left, central and right columns panels, respectively). In this figure,
the projection effects were also explored generating maps where the angle,
$\phi$, between the symmetry axis (the direction along which the ISM has a
gradient in density) and the plane of the sky is 0$\degr$, 45$\degr$, and
60$\degr$ (top, central and bottom row panels, respectively). Overlaying
the X-ray emission, we show contours of the temperature distribution,
integrated along the line of sight.  Because the magnetic field is not
included in our description, it was not possible to generate radio-continuum
maps in order to obtain the actual size of the SNR shock wave.  Therefore, it
was necessary to look for other SNR shock wave tracers. One of them is the
temperature integrated along the line of sight (Rodr\'{\i}guez-Mart\'{\i}nez
et al. 2006) since post-shock temperatures are higher ($T_s \propto v^{2}_s$,
where $v_s$ is the shock-front velocity), reaching values $\sim$ 10$^7$ --
10$^8$ K. Thus, temperature contours describe shock front shapes.

Figure 8 shows that the SNR morphology is practically spherical for model 1,
unlike models 2 and 3, where a pronounced elongation is seen in the horizontal
direction. This effect is more noticeable for model 3 than for model 2. This
morphology is in agreement with radio-continuum images of SNR N~120.
 
With respect to the spatial distribution of the X-ray emission, an increase in
the X-ray surface brightness to the left is observed in all the maps. This
increment is more important in the case of model 3. X-ray observations of this
remnant also show an enhancement in the emission to the east (which
corresponds to the left in our simulations). However, the X-ray maximum is
located inside the SNR shell, close to the eastern border.

To better discriminate between our models and the observed X-ray enhancement
towards the east, we have extracted X-ray surface-brightness profiles
corresponding to the simulated maps displayed in Figure 8, i.e., for each
model and for each viewing angle. The resulting surface-brightness profiles
are displayed in Figure 9. The profiles correspond to horizontal cuts, passing
through the center of the SNR shell. Profiles corresponding to model 3 (H =
3.67 pc) show a stronger enhancement of the X-ray emission to the left
(corresponding to the east) than profiles obtained from model 1 (H = 10 pc).

Table 3 compares the X-ray luminosities obtained for all models, where ISM
absorption associated with a column density of 1.2 $\times$ 10$^{20}$
cm$^{-2}$ has or has not been taken into account. The luminosity for model 3
is 8 $\times$ 10$^{34}$ ergs$^{-1}$, which is in better agreement with the
observed value of 1.2 $\times$ 10$^{35}$ ergs$^{-1}$.
 
A comparison between Figure 9 (simulations) and Figure 6 (observations)
suggests that models 2 and 3 (H = 5 pc and H = 3.67 pc, respectively), and
viewing angles of 45$\degr$ and 60$\degr$, agree with the observed surface
brightness profile in what concerns its actual dimensions (linear diameter
about 30 pc) and slow decrease of X-ray surface brightness from east to west.
On the other hand, model 3 fits best the observed X-ray luminosity of SNR N~120
 (see Table 3).

Figure 10 shows the simulated map corresponding to model 3 (H = 3.67 pc)
viewed with an angle $\phi$=45$\degr$, smoothed with the same angular
resolution as the observations (left panel).  From this map, a
surface-brightness profile was extracted by tracing a cut along the maximum
elongation direction (whose mean position is represented by the gray dashed
line and its width is of 3 pixels in the vertical direction). This profile is
depicted in the right panel of Figure 10 (solid line). A comparison between
Figures 10 and 6 shows that our simulations reproduce the observed difference
in surface brightness between the eastern and the western sides of the SNR, as
well as the global decrease of the X-ray emission towards the west. However,
the observed X-ray brightness profile shows secondary local maxima that are
not reproduced by this model. This difference could arise because SNR N~120 is
evolving in an ISM which has, in addition to the global density gradient in
the east-west direction, several clumps or inhomogeneities. If one of these
clumps was recently swept up by the remnant blast wave, its effect will be an
increase of the X-ray emission projected onto the SNR interior.

To determine the importance of thermal conduction in the evolution of the SNR,
another simulation, model 4, was made. This simulation employs the same
parameters as model 3, with the inclusion of thermal conduction. The heat flux
thermal conduction is included in the YGUAZ\'U code by employing Spitzer's
(1962) classical expression $q= \kappa \nabla T$, where $\kappa= \beta
T^{2.5}$ ( with $\beta \simeq 6 \times 10^{-7}erg~s^{-1}K^{-1}cm^{-1}$), when
the electron collisional mean free path $\lambda_{e}$ is less than the
temperature scale height $L_{T}=T/ {|\nabla T|}$. And, $q_{s}=-5 \phi_{s}cP$
(Cowie \& McKee 1977), where c is the isothermal sound speed, P is the gas
pressure, and $\phi_{s}$ is a dimensionless parameter of the order of 1, when
$\lambda_{e} > L_{T}$ and the heat flux is saturated. For details on the
inclusion of classical and saturated thermal conduction in the YGUAZ\'U-A code
see Vel\'azquez et al.(2004).

The luminosities found in the models with and without thermal conduction
differ by only $10\%$. In Figure 10 we also present the surface brightness
distribution along the maximum elongation predicted by model 4 (dashed lines).
We have superposed the radial surface brightness profiles with and without
considering thermal conduction. As seen in this figure, the effects of thermal
conduction on the X-ray emission of SNR N~120 are negligible, as expected.
Previous works have reported that the effects of thermal conduction are
important for the case of more evolved SNRs, which have entered the third
phase of evolution or radiative phase (Tilley et al. 2006). Therefore, the
essential parameter in the evolution of the SNR is the density gradient of the
interstellar medium.

We conclude that an exponential density gradient with a scale height of H =
3.67 pc for the ambient medium, where the SNR evolves, and view with an angle
between 45$\degr$ and 60$\degr$, reproduce the main observed issues: total
X-ray luminosity, dimensions and peculiar X-ray surface brightness profile.
From the results of the last run (model 4) we also conclude that thermal
conduction effects are negligible.

\section{Summary}

In this work we have presented XMM-Newton images and spectra of SNR N~120 in
the LMC together with previous optical kinematic data, as well as numerical
simulations that reproduce quite well the X-ray luminosity and surface
brightness distribution of the SNR. Furthermore, the physical size and 
shock wave expansion velocity obtained from our numerical simulations are in
agreement with the obtained ones by a previous kinematical study by Rosado et
al. (1993).

We have found from XMM-Newton data that the X-ray emission of SNR N~120
consists of a conspicuous elliptical patch elongated along the E-W direction
in the same way as the radio-continuum emission and with similar extension.

We have also found that the X-ray emission shows an asymmetry in brightness
along the E-W direction and we have quantified this asymmetry by means of
surface brightness profiles showing that the maximum X-ray surface brightness
is emitted from the eastern boundary of the SNR. Secondary peaks in the
brightness profile are also detected.

We have fitted reasonably well the X-ray spectrum of SNR N~120 with that of
the thermal gas with LMC chemical abundances. Hence, we can conclude that the
soft X-ray spectrum of this SNR is thermal.

We have also detected some faint and diffuse X-ray emission which shows some
spatial correlation with the H$\alpha$ emission of the N~120 superbubble where
SNR N~120 is at its boundaries. In view of the correlation, we suggest that
the faint emission comes from the N~120 superbubble.

Working with the hypothesis that the asymmetry in X-ray emission is produced
by the expansion of an SNR in an ISM with an exponential density profile, we
have carried out several 2D axisymmetric numerical simulations of this model
in order to check the validity of that hypothesis.

The density gradient could be due to the particular environment where the SNR
N~120 is located: forming part of the boundary of the large nebular complex
N~120.

The model that best fits the X-ray observations is the one with a density
decreasing exponentially from the N~120 superbubble, with a scale length of
3.7 pc and with the assumption that the elongated X-ray emission is seen with
its major axis inclined between 45$\degr$ and 60$\degr$ with respect to the
plane of the sky. Secondary peaks in the X-ray surface brightness profile are
not reproduced in our simulations and are probably due to the existence of ISM
clumps recently shocked by the SNR blast wave.

We have also shown, by means of another numerical simulation (model 4), that
thermal conduction effects are negligible for this SNR.

\acknowledgments Authors acknowledge anonymous referee for her/his very useful
comments which allow us to improve this manuscript. We also acknowledge Rosa
Murphy-Williams and You Hua Chu for their comments and suggestions. 
The authors thank Dr. Elena
Jim\'enez for her useful comments and advice during the development of this
article. We thank Liliana Hern\'andez, Alfredo D\'\i az, Francisco Ruiz and
Carmelo Guzm\'an (IA-UNAM)for computer help, Enrique Palacios, Mart\'\i n Cruz
and Antonio Ram\'\i rez (ICN-UNAM) for maintaining our linux servers where the
numerical simulations were carried out, and Stan Kurtz (CRyA, UNAM)
for reading the
manuscript.  We also thank Sean Points and the MCELS collaboration for the use
of the MCELS images.  This paper was done with financial support from grants
46054-F, 46828-F, 40095-F from CONACYT, IN108207 and IN100606 from DGAPA-UNAM.

\clearpage

%% the file, uncomment the \plotone, \plottwo, and
%% If you need a layout that cannot be achieved with \plotone or
%% \plottwo, you can invoke the graphicx package directly with the
%% \includegraphics command or use \plotfiddle. For more information,
%% please see the tutorial on "Using Electronic Art with AASTeX" in the
%% documentation section at the AASTeX Web site,
%% http://www.journals.uchicago.edu/AAS/AASTeX.
%%
%% The examples below also include sample markup for submission of
%% supplemental electronic materials. As always, be sure to check
%% the instructions to authors for the journal you are submitting to
%% for specific submissions guidelines as they vary from

%% This example uses \plotone to include an EPS file scaled to
%% 80% of its natural size with \epsscale. Its caption
%% has been written to indicate that additional figure parts will be
%% available in the electronic journal.

\begin{figure}
\epsscale{1.0}
\plotone{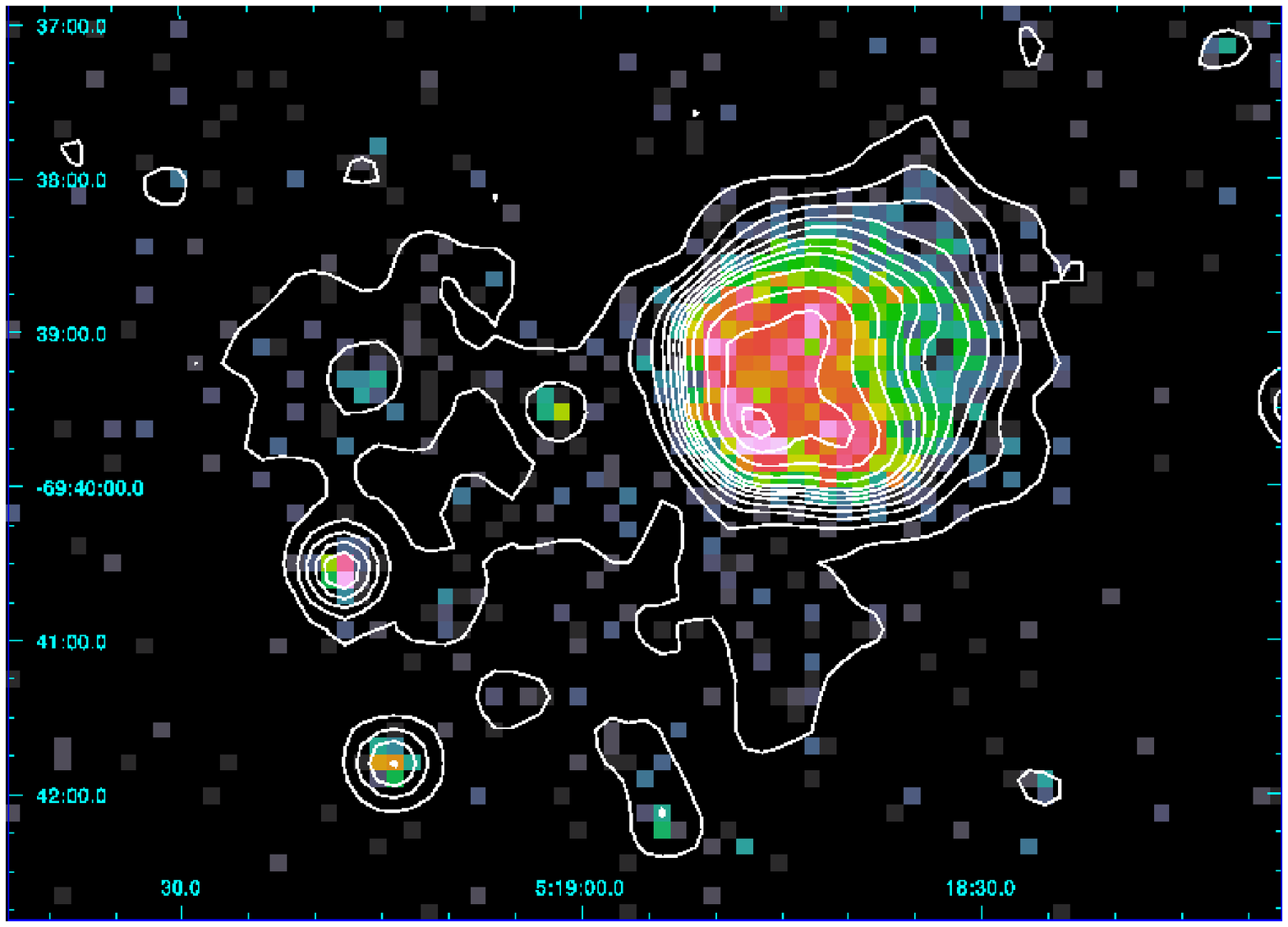}
\caption{EPIC/MOS1$+$MOS2 color image of the X-ray emission of the 
SNR N~120 in  the energy band from 0.2 to 2.0 keV. In white are overplotted the X-ray 
contours of the same X-ray image revealing the shape of the X-ray emission (from the 3rd contour). 
Contours have been drawn at levels of 1$\sigma$, 2$\sigma$, 3$\sigma$, 4$\sigma$, 5$\sigma$, 6$\sigma$, 7$\sigma$, 8$\sigma$, 10$\sigma$, 13$\sigma$, 16$\sigma$ and 18$\sigma$ above the background level.}
\label{figure 1}
\end{figure}

\clearpage

\begin{figure}
\begin{center}
\includegraphics[width=1.0\textwidth]{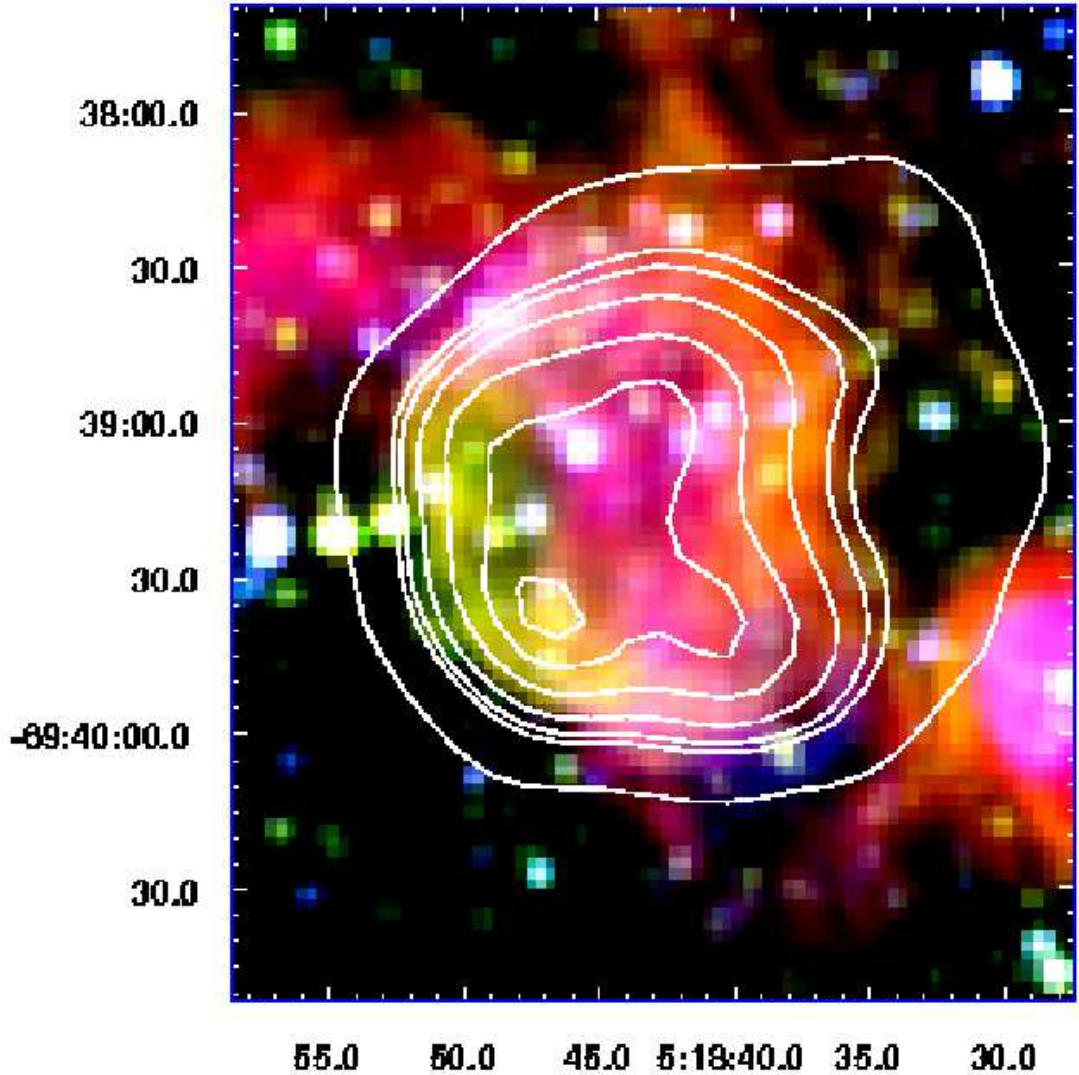}

\caption{Three-color picture  made up of Magellanic Clouds Emission-Line Survey (MCELS) images in H$\alpha$ (red), [SII] (green), and [OIII] (blue) of the SNR N~120. Superposed onto this picture are some of the X-ray contours (specially at levels of 3$\sigma$, 7$\sigma$, 8$\sigma$, 10$\sigma$, 13$\sigma$, 16$\sigma$ and 18$\sigma$) displayed in Figure 1. }
\label{fig2}
\end{center}
\end{figure}

\clearpage

\begin{figure}
\epsscale{1}
\plotone{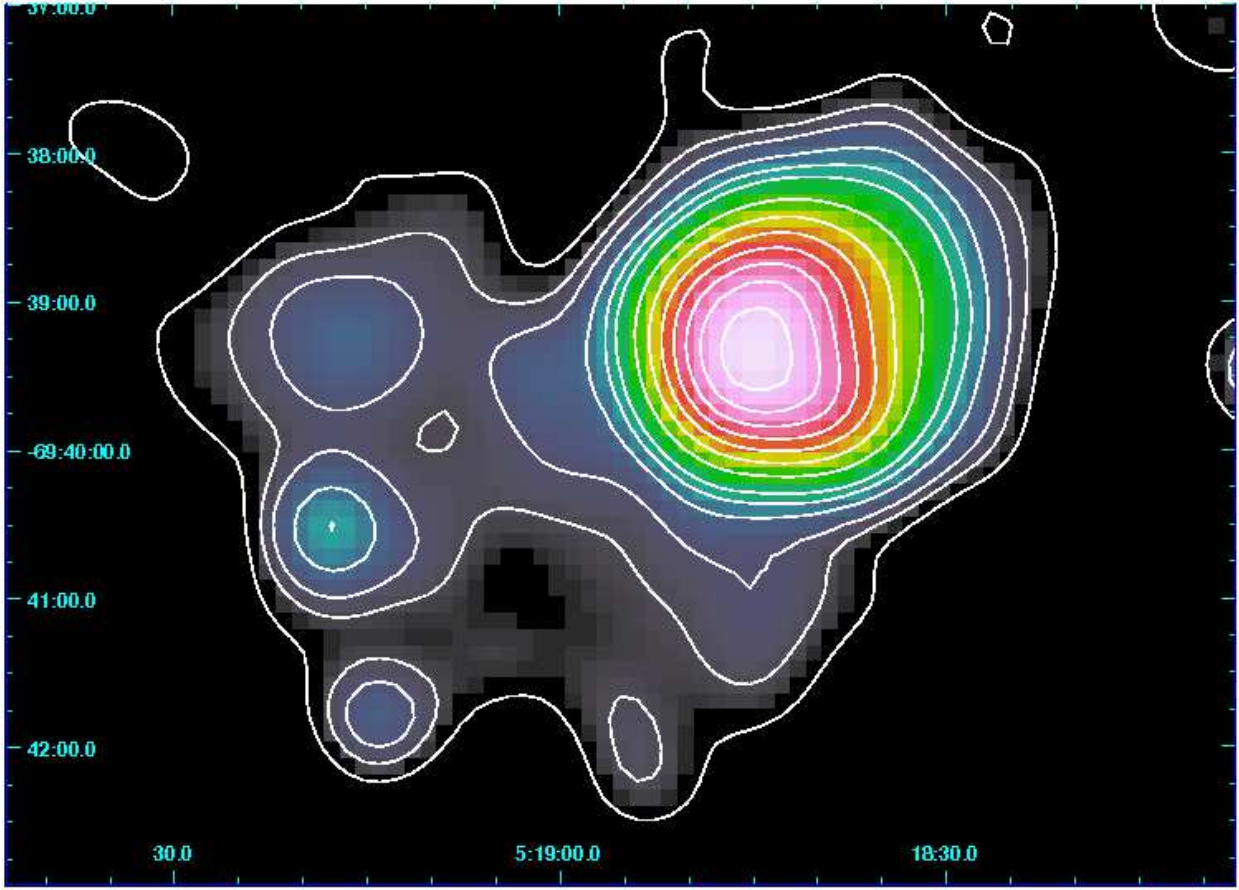}
\caption{ EPIC/MOS1$+$MOS2 Gaussian smoothed color image of the X-ray emission of the 
SNR N~120 in  the energy band from 0.2 to 2.0 keV. In white are overplotted the X-ray 
contours of the same X-ray image. 
Contours have been drawn at levels of 4$\sigma$, 6$\sigma$, 8$\sigma$, 12$\sigma$, 16$\sigma$, 20$\sigma$, 30$\sigma$, 40$\sigma$, 50$\sigma$, 60$\sigma$, 70$\sigma$, 80$\sigma$, 90$\sigma$ and 100$\sigma$ above the background level. }
%%\caption{Three-color picture  made up of Magellanic Clouds Emission-Line Survey images in H$\alpha$ (red), [SII] (green), and [OIII] (blue) of the N~120 nebular complex in the LMC. SNR N~120 is seen as the elongated bubble to the West. Superposed onto this picture are the X-ray contours of the 0.2-2.0 keV band displayed in Figure 2, in order to show the location of the diffuse X-ray emission. The optical images have $2\arcsec\times2\arcsec$ pixels spatial resolution.}
\label{fig3}
\end{figure}

\clearpage

\begin{figure}
\epsscale{1}
\plotone{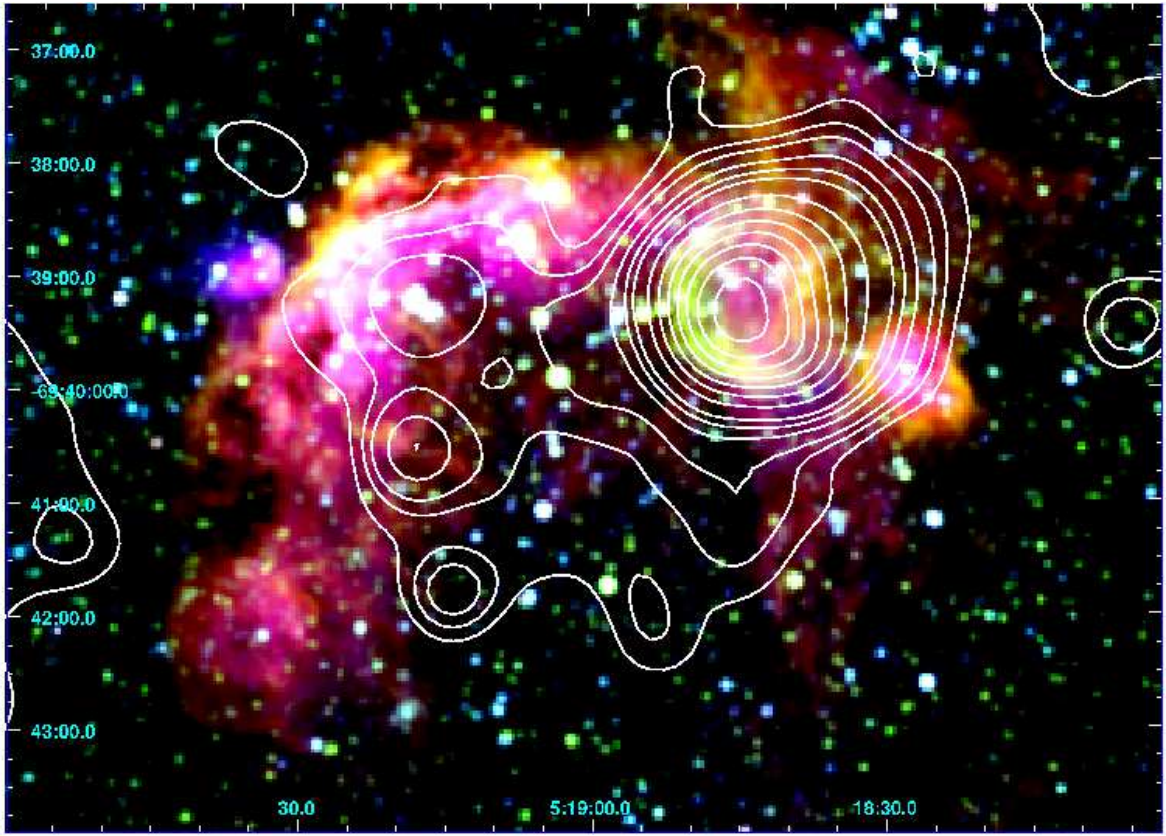}
\caption{Three-color picture  made up of MCELS images in H$\alpha$ (red), [SII] (green), and [OIII] (blue) of the N~120 nebular complex
in the LMC. SNR N~120 is seen as the elongated bubble to the West. Superposed onto this picture are the X-ray contours of the 0.2-2.0 keV band displayed in Figure 3, in order to show the location of the diffuse X-ray emission.
The optical images have $2\arcsec\times2\arcsec$ pixels spatial resolution.}
%%\caption{Close-up of figure 3. Superposed onto this picture are some of the X-ray contours (specially at levels of 3$\sigma$, 7$\sigma$, 8$\sigma$, 10$\sigma$, 13$\sigma$, 16$\sigma$ and 18$\sigma$) displayed in figure 1, to appreciate better the emission in the optical.}
\label{fig4}
\end{figure}

\clearpage
%% Here we use \plottwo to present two versions of the same figure,
%% one in black and white for print the other in RGB color
%% for online presentation. Note that the caption indicates
%% that a color version of the figure will be available online.
%%

%%\begin{figure}
%%\plottwo{jorge1.ps}{jorgecontours1.ps}
%%\caption{A panel taken from Figure 2 of \citet{rudnick03}. 
%%See the electronic edition of the Journal for a color version 
%%of this figure.\label{fig2}}
%%\end{figure}
%%\clearpage

\begin{figure}
\epsscale{1.0}
\plotone{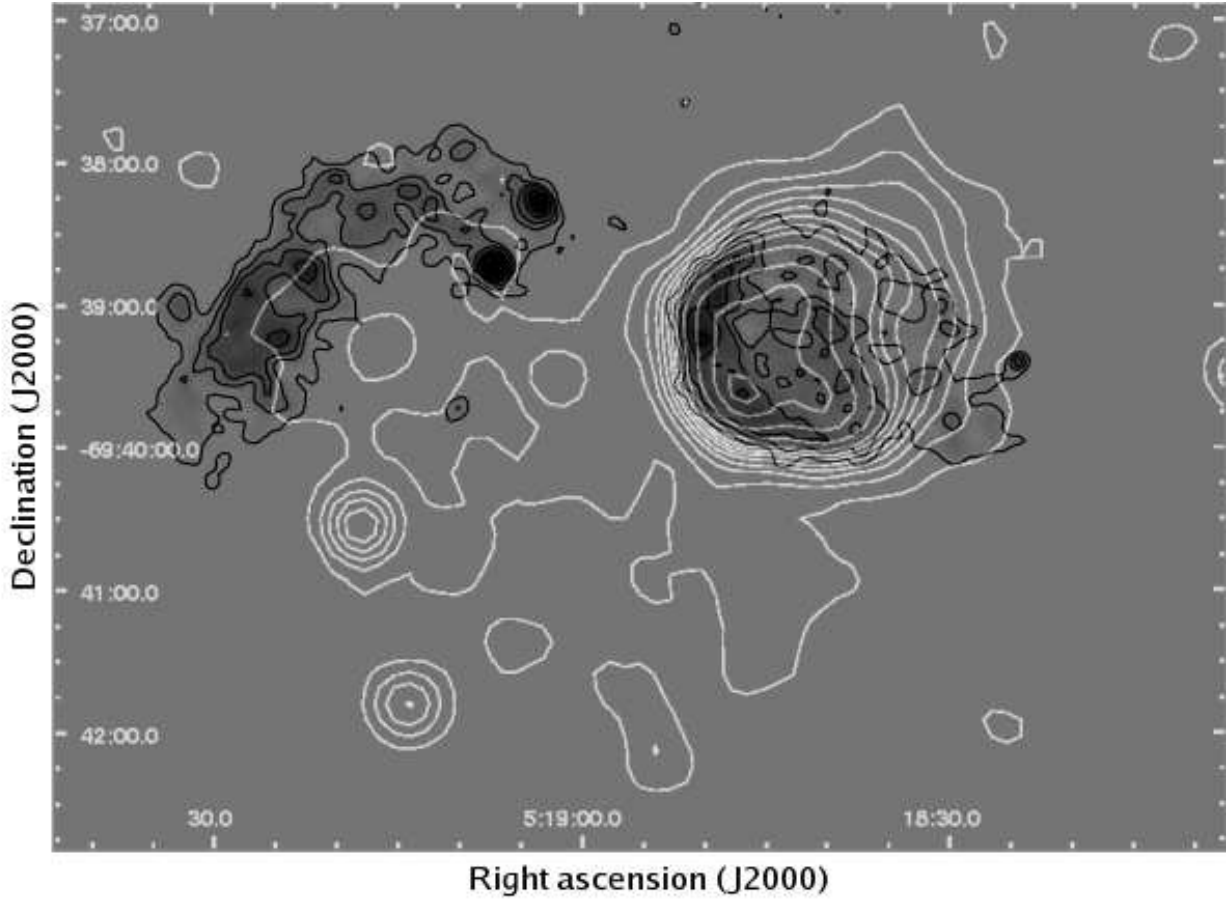}
\caption{Radio contours of equal surface-brightness (in black) taken from Dickel \& Milne (1998). Superposed on the radio contours are the X-ray contours of the 0.2-2.0 keV band  (in white) displayed in Figure 1 (i.e., without any spatial smoothing ).  }
\label{fig5}
\end{figure}
\clearpage

\begin{figure}
\epsscale{1.10}
\plotone{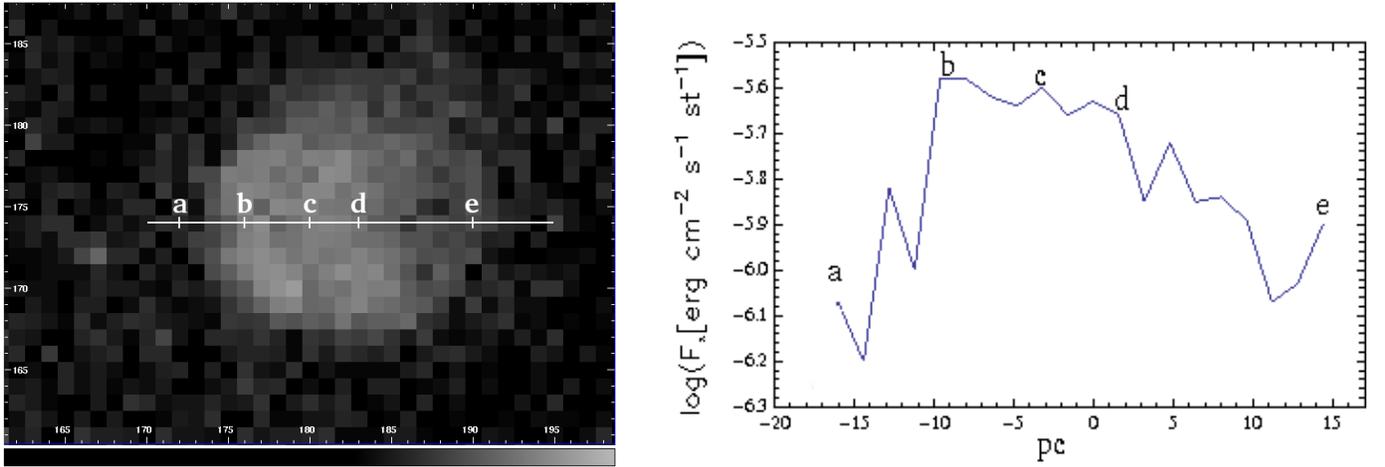}
\caption{Left panel: EPIC/MOS X-ray emission map for SNR N~120 in the LMC. The white, horizontal line corresponds to the mean position of the `slit' (the width of the aperture is $19\arcsec $ (5 pc) and the length is 174$\arcsec$ (42 pc)) along which the brightness profile was extracted. Right panel: X-ray surface brightness profile extracted along the horizontal line shown in the left panel. Some marks in both panels are given as reference of different locations.}
\label{fig6}
\end{figure}
\clearpage

%% This figure uses \includegraphics to scale and rotate the still frame

\newpage

\begin{figure}
\begin{center}
%\centering
\includegraphics[width=1.0\textwidth]{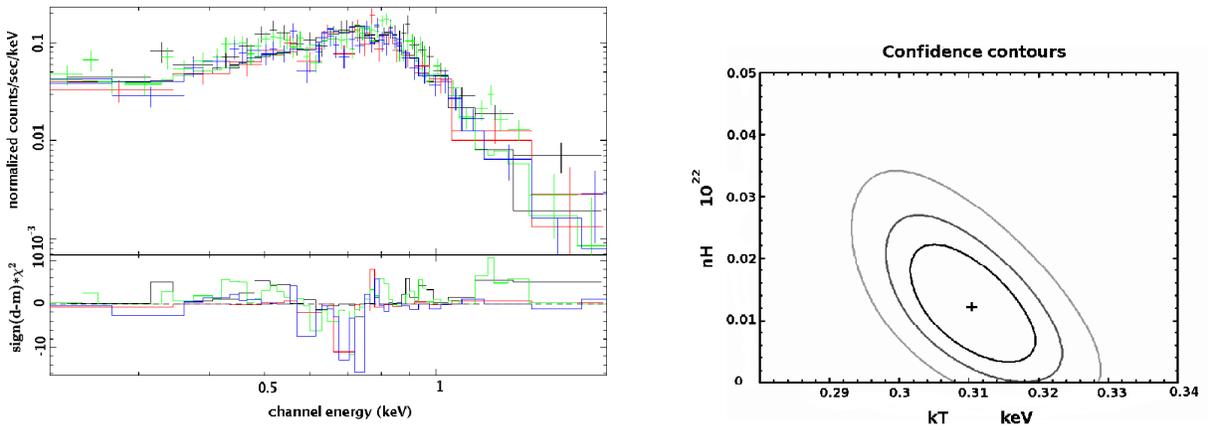}
\caption{Left panel: X-ray spectrum of the scheduled MOS1 and MOS2 data (in black and red, respectively), and of unscheduled MOS1 and MOS2 data (in green and blue, respectively) analized jointly. Right panel:$\Delta\chi^2$ contours for an absorbed MEKAL: contours for 2D confidence levels of 68$\%$ (internal contour), 90$\%$ and 99$\%$ (external contour) are drawn. The cross 
indicates the model used in the fit of the left panel.}
\label{fig7}
\end{center}
\end{figure}

\newpage

\begin{figure}
\epsscale{0.8}
\plotone{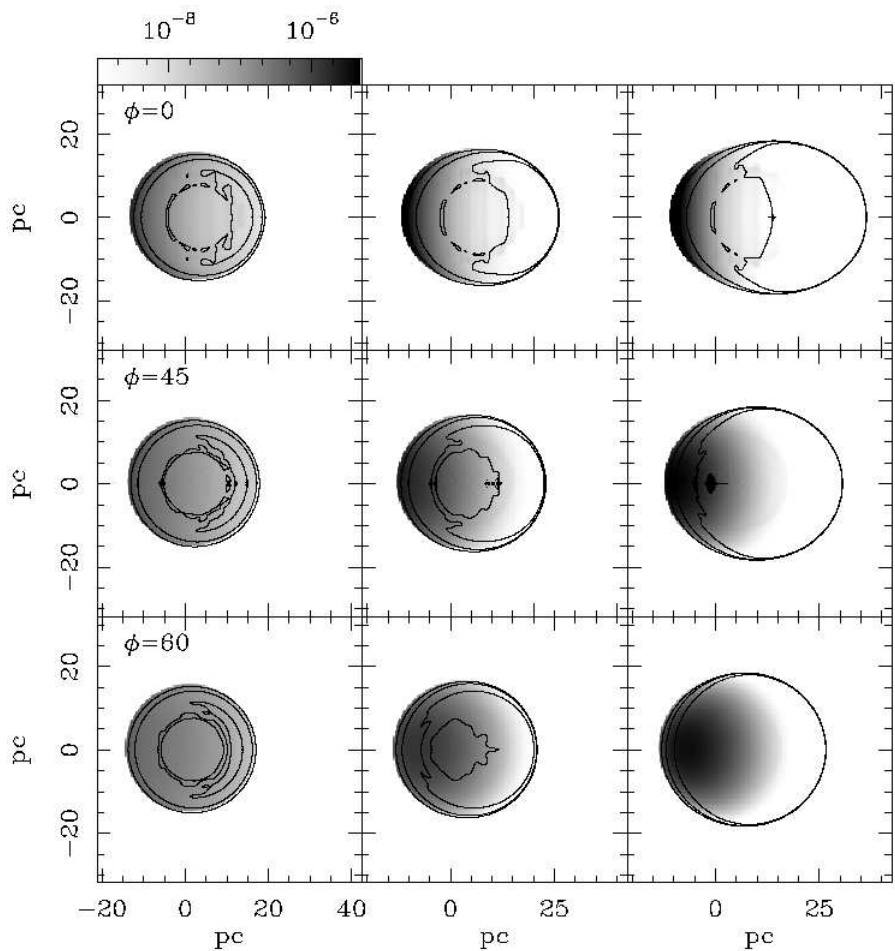}
\caption{Simulated X-ray emission maps for models 1, 2 and 3 (left, central and  right column panels, respectively). Different viewing angles between the symmetry axis and the plane of the sky, $\phi$ = 0$\degr$, 45$\degr$ and 60$\degr$, are considered (top, central and bottom row panels, respectively). Overlaying the X-ray emission are plotted some contours of the temperature distribution, which trace the shock wave position. 
The linear grey-scale is given in units ergs cm$^{-2}$ s$^{-1}$  sr$^{-1}$;
vertical and horizontal axis are given in pc.}
\label{fig8}
\end{figure}
\clearpage

\begin{figure}
\epsscale{0.5}
\plotone{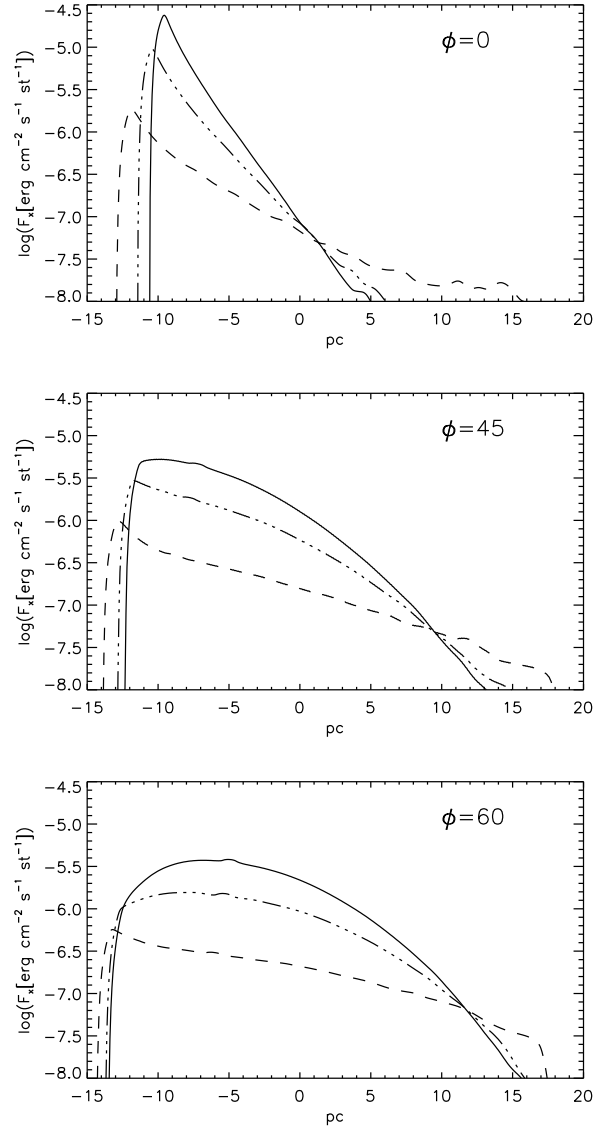}
\caption{X-ray surface-brightness profiles for all panels shown in Figure 8. These profiles correspond to horizontal cuts, passing by the center of the SNR shell. The different viewing angles are marked in each panel. Solid lines: model 3, point hatched lines: model 2, and hatched lines: model 1.}
\label{fig9}
\end{figure}
\clearpage

\begin{figure}
\epsscale{1.0}
\plotone{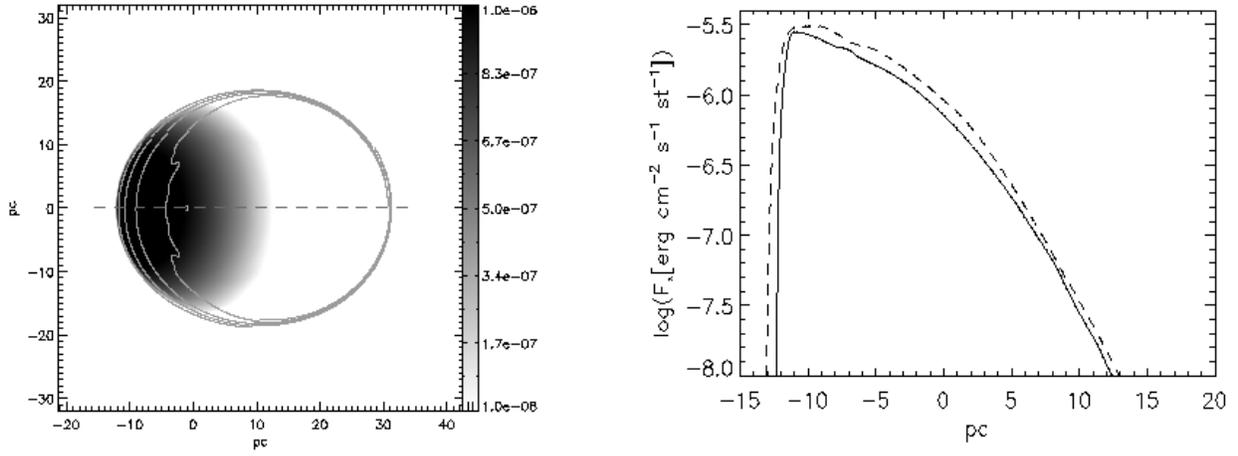}
\caption{Left panel: simulated X-ray emission map for model 3 (H = 3.7 pc) and viewing angle $\phi$ = 45$\degr$, smoothed at the same angular resolution than the observations. The dashed, horizontal line corresponds to the position along which the brightness profile was extracted. Right panel: X-ray surface-brightness profiles extracted along the horizontal of the left panel. 
The profile obtained for the case where thermal conduction is included in the calculation is displayed in dashed lines, while the case where thermal conduction is not included is represented by a solid line.}
\label{fig10}
\end{figure}
\clearpage

\renewcommand{\arraystretch}{0.7}
\begin{deluxetable}{lccc}
\tabletypesize{\scriptsize}
\tablecaption{Parameters of Different Models Fitted to X-ray Spectra \label{data}}
\tablewidth{0pt}
\tablehead{ Parameter
& \colhead{MEKAL\tablenotemark{a}} & \colhead{PSHOCK} & 
\colhead{NEI}\\
& & & }
\startdata
N$_{H}$ & 1.21 $\pm$0.66 & 2.0 & 2.0 \\
 
 (10$^{20} cm^{-2}$)& & (frozen) & (frozen)\\
& & &  \\
kT & 0.31 keV & 0.49 keV & 0.95 keV \\
& & &  \\

Reduced $\chi^2$ & 1.60 & 1.35 & 1.93 \\
& & &  \\

$\tau$ & -- & 0.0-- 5.64 $\times 10^{11}$ &  2.32 $\times 10^{10}$\\
(cm$^{-3}$s)& & &  \\

\enddata
\tablenotetext{a}{ Average LMC abundances = 0.3 Solar abundances}
\end{deluxetable}
\renewcommand{\arraystretch}{1}
\clearpage

\begin{table}
\begin{center}
\caption{MEKAL parameters corresponding to a simultaneous fit of the MOS1 and MOS2, scheduled and unscheduled data.}
%%\begin{tabular}{crr}
\begin{tabular}{|l|l|c|}\hline
\tableline\tableline

\multicolumn{1}{c}{Parameters}\\
%\midrule
\tableline

 N$_H$ $( 10^{20}$ cm$^{-2})$ & $1.2\pm0.66$ \\ \hline
   
   kT (keV)&$0.31 \pm 0.06$ \\ \hline
   He & 1.0 \tablenotemark{a} \\ \hline
   C & 0.3 \tablenotemark{a} \\ \hline
   N & 0.3 \tablenotemark{a} \\ \hline
   O & 0.3 \tablenotemark{a}\\ \hline
  Ne & 0.3 \tablenotemark{a}\\ \hline
  Na & 0.3 \tablenotemark{a}\\ \hline
  Mg & 0.3 \tablenotemark{a}\\ \hline
  Al & 0.3 \tablenotemark{a}\\ \hline
  Si & 0.3 \tablenotemark{a}\\ \hline
   S & 0.3 \tablenotemark{a}\\ \hline
   Ar & 0.3 \tablenotemark{a}\\ \hline
   Ca & 0.3 \tablenotemark{a}\\ \hline
   Fe & 0.3 \tablenotemark{a}\\ \hline
   Ni & 0.3 \tablenotemark{a}\\ \hline
   $L_x$ $(10^{35}\ \rm{ergs\ s^{-1}})$   & 1.2 \\ \hline
   Flux $(10^{-13}\ \rm{ergs\ cm^{-2} s^{-1}})$ &4.17 \\ \hline

\tableline
\end{tabular}
%% Any table notes must follow the \end{tabular} command.
\tablenotetext{a}{ Average LMC abundances}
%%\tablenotetext{b}{Yet another sample footnote for table~\ref{tbl-2}}
%%\tablenotetext{c}{Another sample footnote for table~\ref{tbl-2}}
%%\tablecomments{We can also attach a long-ish paragraph of explanatory
%%material to a table.}
\end{center}
\end{table}
\clearpage

\renewcommand{\arraystretch}{0.7}
\begin{deluxetable}{lcrrr}
\tabletypesize{\scriptsize}
\tablecaption{X-ray luminosities from different models \label{data}}
\tablewidth{0pt}
\tablehead{
& \colhead{model 1} & \colhead{model 2} & 
\colhead{model 3}& model 4 (with\\
& & & &thermal conduction)}
\startdata
Length scale & 10 pc & 5 pc & 3.67 pc & 3.67 pc\\
 
 & & & & \\
$L_{x}(10^{35} erg s^{-1})$ & 0.16 & 0.44 & 0.81& \hspace{0.8cm} 0.92 \\
 
 & & & & \\
$L_{x,abs}(10^{35} erg s^{-1})$ & 0.13&0.34&0.60&\hspace{0.8cm} 0.72\\

\enddata
\end{deluxetable}
\renewcommand{\arraystretch}{1}

%% If the table is more than one page long, the width of the table can vary
%% from page to page when the default \tablewidth is used, as below.  The
%% individual table widths for each page will be written to the log file; a
%% maximum tablewidth for the table can be computed from these values.
%% The \tablewidth argument can then be reset and the file reprocessed, so
%% that the table is of uniform width throughout. Try getting the widths
%% from the log file and changing the \tablewidth parameter to see how
%% adjusting this value affects table formatting.

%% The \dataset{} macro has also been applied to a few of the objects to
%% show how many observations can be tagged in a table.

%\clearpage

%% Tables may also be prepared as separate files. See the accompanying
%% sample file table.tex for an example of an external table file.
%% To include an external file in your main document, use the \input
%% command. Uncomment the line below to include table.tex in this
%% sample file. (Note that you will need to comment out the \documentclass,
%% \begin{document}, and \end{document} commands from table.tex if you want
%% to include it in this document.)


%% The following command ends your manuscript. LaTeX will ignore any text

\end{document}